\documentclass[iop,twocolumn,numberedappendix,twocolappendix]{emulateapj}

\newcommand{\cxo}{{\em Chandra}}

\newcommand{\lsim}{\raisebox{-.4ex}{$\stackrel{<}{\scriptstyle \sim}$}}

\newcommand{\mim}{\mbox{$\mu$m}}

\newcommand{\ngc}{\mbox{NGC\,602a}}
\newcounter{ion} 

\def \etal   {\hbox{et~al.\/}}
\def\changed{}

\shorttitle{X-rays from young suns in the SMC}
\shortauthors{Oskinova et al.}

\begin{document}
\title{Discovery of X-ray emission from young suns in the Small Magellanic Cloud}

\author{L. M. Oskinova\altaffilmark{1}
W. Sun\altaffilmark{2},
C. J. Evans\altaffilmark{3},
V. H\'enault-Brunet\altaffilmark{4},
Y.-H. Chu\altaffilmark{5},\\
J. S. Gallagher III\altaffilmark{6},
M. A. Guerrero\altaffilmark{7}
R. A. Gruendl\altaffilmark{5},
M. G\"udel\altaffilmark{8},
S. Silich\altaffilmark{9},
Y. Chen\altaffilmark{2},\\
Y. Naz\'e\altaffilmark{10},
R. Hainich\altaffilmark{1},
J. Reyes-Iturbide\altaffilmark{11}}

\altaffiltext{1}{Institute for Physics and Astronomy, University Potsdam, 14476 Potsdam, Germany}
\altaffiltext{2}{Department of Astronomy, Nanjing University, Nanjing, 210093 Jiangsu, China}
\altaffiltext{3}{UK Astronomy Technology Centre, Royal Observatory Edinburgh, Blackford Hill, 
Edinburgh, EH9 3HJ, UK}
\altaffiltext{4}{Scottish Universities Physics Alliance (SUPA), Institute for Astronomy, 
University of Edinburgh, Blackford Hill, Edinburgh EH9 3HJ, UK}
\altaffiltext{5}{Department of Astronomy, University of Illinois, 1002 West Green Street, 
Urbana, IL 61801, USA}
\altaffiltext{6}{Department of Astronomy, University of Wisconsin-Madison, 5534 Sterling, 475 North 
Charter Street, Madison, WI 53706, USA}
\altaffiltext{7}
{Instituto de Astrof\'{\i}sica de Andaluc\'{\i}a, IAA-CSIC, Glorieta de 
  la Astronom\'ia s/n, 18008 Granada, Spain}
\altaffiltext{8}{University of Vienna, Department of Astrophysics, T\"urkenschanzstrasse 17, 
1180, Vienna, Austria}
\altaffiltext{9}{Instituto Nacional de Astrof\'{\i}sica Optica y Electr\'onica, AP 51, 
72000 Puebla, Mexico}
\altaffiltext{10}{GAPHE, D\'{e}partement AGO, Universit\'{e} de Li\`{e}ge, All\'{e}e du 6 Ao\^{u}t 17, 
Bat. B5C, B4000 Li\`{e}ge, Belgium}
\altaffiltext{11}{LATO-DCET/Universidade Estadual de Santa Cruz, 
Rodovia Jorge Amado, km
16, 45662-000 Ilh\'eus, BA, Brazil}

\begin{abstract}    
We report the discovery of extended X-ray emission within the young star
cluster \ngc\ in the Wing of the Small Magellanic Cloud (SMC) based on
observations obtained with the \cxo\ X-ray Observatory.  X-ray emission is
detected from the cluster core area with the highest stellar density and
from a dusty ridge surrounding the H\,{\sc ii} region.  We use a census
of massive stars in the cluster to demonstrate that a cluster wind or
wind-blown bubble is unlikely to provide a significant contribution to the
X-ray emission detected from the central area of the cluster. We therefore
suggest that X-ray emission at the cluster core originates from an
ensemble of low- and solar-mass pre-main-sequence (PMS) stars, each of
which would be too weak in X-rays to be detected individually. We
attribute the X-ray emission from the dusty ridge to the embedded tight
cluster of the new-born stars known in this area from infrared
studies. Assuming that the levels of X-ray activity in young stars in the
low-metallicity environment of NGC\,602a are comparable to their Galactic
counterparts, then the detected spatial distribution, spectral properties,
and level of X-ray emission are  largely consistent with those expected 
from low- and solar-mass PMS stars and young stellar objects (YSOs). This 
is the first discovery of X-ray emission attributable to PMS stars and YSOs 
in the SMC, which suggests that the accretion and dynamo processes in young, low-mass
objects in the SMC resemble those in the Galaxy.
\end{abstract}



\section{Introduction}

The eastern `Wing' of the Small Magellanic Cloud (SMC) provides us
with an excellent laboratory to investigate the role of environment in star
formation and stellar evolution when compared to Galactic studies.  The Wing 
has a low content of gas, dust and stars, with a comparably low
metallicity to that found in the main body of the SMC
\citep[e.g.,][]{Lee2005}. These are typical conditions for
low-metallicity dwarf irregular galaxies, which are the most common
type among {\em all} star-forming galaxies \citep{Gallagher1984}.

The most significant site of star formation in the Wing is NGC\,602,
which is a conglomerate of at least three stellar clusters: NGC\,602a
\citep[immersed in the LHA\,115-N\,90 H\,{\sc ii}
region,][]{Henize1956}, with NGC\,602b adjacent to the north, and
NGC\,602c $\sim$$11'$ to the northeast \citep{West1964}.  
\citet{Cignoni2009} advocated a distance modulus to the young 
stellar population of NGC\,602a of 18.7\,mag 
\citep[also see the discussion by][]{Evans2012}; we adopt
this distance in the analysis presented here.

\begin{figure*}
\centering
\includegraphics[width=1.9\columnwidth]{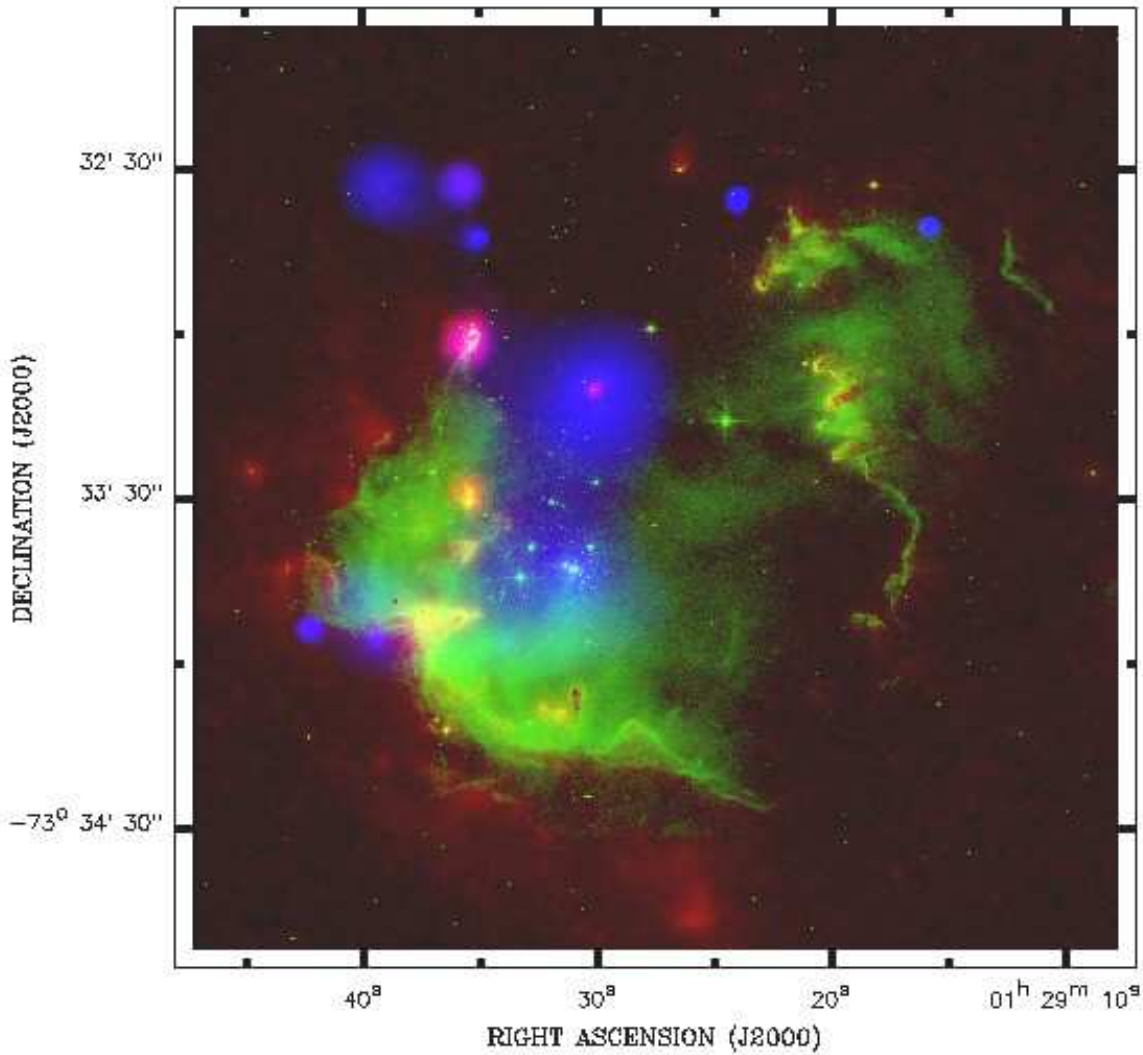}
\caption{Color-composite image of \ngc, constructed from observations
  with {\em Spitzer}--IRAC at 8\,\mim\ (red, AOR\,12485120), {\em HST}--ACS with the
  F658N filter (green, data set J92FA6), and \cxo\ ACIS-I in the 0.5-7.0\,keV band
  (adaptively smoothed, in blue). Image size is $\approx 2'\times 2'$ 
($\approx 32$\,pc$\times$32\,pc).}
\label{fig:combi} 
\end{figure*}

The images
of \ngc\ from the {\em Hubble Space Telescope (HST)} Advanced Camera
for Surveys
(ACS)\footnote{http://hubblesite.org/newscenter/archive/releases/2007/04}
reveal a star-forming region  with a striking ring morphology, as
shown in Fig.\,\ref{fig:combi}.  Massive OB-type stars shine within
the broken ring, while lower-mass, pre-main-sequence (PMS) stars are
distributed around them \citep[e.g.][]{Carlson2007, Goul2007,
  Goul2012}.  Infrared (IR) images from the {\em Spitzer Space
  Telescope} show the same morphology, with numerous embedded, young
stellar objects (YSOs) revealed in the dusty ridges
\citep[e.g.][]{Carlson2011}.

From analysis of the {\em HST}-ACS and {\em Spitzer} observations,
\citet{Cignoni2009} and \citet{Carlson2011} have argued that the stars
in \ngc\ belong to one of three distinct age groups: (i) 6-8\,Gyr old
very metal-poor field stars; (ii) hot massive stars with ages of a
few Myr responsible for the ionization of N\,90 and low-mass PMS
stars of the same age; (iii) tens of kyr old YSOs (Class 0.5-I)
embedded in the dusty ridges and pillars.  The cluster mass has been
estimated as $\sim 2000$\,$M_\odot$ 
\citep{Cignoni2009, Carlson2011}.  The
star-formation rate in NGC\,602a was determined by \citet{Cignoni2009}
to have reached $(0.3-0.7)\times 10^{-3}$\,$M_\odot$\,yr$^{-1}$ in the
last 2.5\,Myr, comparable to that found in Galactic OB associations.

As part of a combined X-ray--optical study of NGC\,602 and its
associated H$\alpha$ supergiant shell
\citep[`SGS-SMC1',][]{Meaburn1980} we obtained deep X-ray imaging of
NGC\,602 with the {\em Chandra X-ray Observatory}, giving the best
opportunity to date to study the SMC's Wing at X-ray wavelengths.
This paper presents and discusses these new observations.

In the context of deep X-ray observations, the Orion Nebular Cluster
(ONC, M\,42) provides a useful Galactic comparison.    Indeed, 
the ONC -- located
at a distance of $\sim$0.4\,kpc \citep[e.g.][]{Sandstrom2007} -- is
one of the best-studied young, massive star clusters. It has a cluster
stellar mass of $\approx$\,1800\,$M_\odot$ within
2\,pc \citep{Hillenbrand1998}\footnote{Note that \citet{Alves2012} found
a rich stellar population in front of the Orion~A cloud, in which the
ONC is embedded. This 4-5\,Myr old cluster (NGC\,1980) in front of the 
ONC is quite massive ($\sim$1\,000$M_\odot$) and overlaps significantly 
with what is usually assumed to be the ONC. It could account for more
than 10-20\% of what is taken in the literature as the ONC population.}
 and contains multiple stellar generations
with ages of $0.1-3$\,Myr, with a lower limit to the recent
star-formation rate of $\sim 10^{-4}\,M_\odot$\,yr$^{-1}$.  The
cluster core is compact with a radius $\lsim 0.5$\,pc
\citep{Hillenbrand1997, Hillenbrand1998}.  Most pertinently, the ONC
has been studied extensively in X-rays \citep[e.g.,][]{Schulz2001,
  Feigelson2002, Flaccomio2003, Gudel2008}, with deep \cxo\
observations revealing the X-ray properties of its young stars
\citep[see the ApJS special issue, vol.\,160,][]{Getman2005}, which
are considered a template for massive star-forming region.

The mass, size, range in stellar ages, and star-formation history of 
the ONC are comparable to NGC\,602a. 
The stellar initial mass function (IMF) for stars more massive than
1\,$M_\odot$ in NGC\,602a was found by \citet{Schmalzl2008} to be
similar to a Salpeter IMF, i.e., comparable to the field IMF in the
solar neighbourhood.  
Albeit there is a difference in the exact number
  of high-mass stars between the ONC \citep[6
  stars,][]{Hillenbrand1997} vs. NGC\,602a \citep[10
  stars,][]{Schmalzl2008}, such differences are to be expected
  from stochastic sampling of the IMF in relatively low-mass star
  clusters \citep[e.g.][]{Cervino2003}.

\begin{deluxetable*}{lcccccrr}
  \tabletypesize{\footnotesize}
  \tablecaption{{\em Chandra} X-ray Point Sources in \ngc \label{tab:ngc602src}}
  \tablewidth{0pt}
  \tablehead{
  \colhead{Src} &
  \colhead{CXOU Name} &
  \colhead{$\delta_x$ ($''$)} &
  \colhead{CR (cnt\,ks$^{-1})$} &
  \colhead{HR} &
  \colhead{HR1} &
  \colhead{Flag} &
  \colhead{Counterparts} \\
  \noalign{\smallskip}
  \colhead{(1)} &
  \colhead{(2)} &
  \colhead{(3)} &
  \colhead{(4)} &
  \colhead{(5)} &
  \colhead{(6)} &
  \colhead{(7)} &
  \colhead{(8)} 
  }
\startdata
1 &  J012915.85-733240.7 & 0.7 &$\phantom{1}0.05 \pm 0.02$& --& --& B & ? \\
2 &  J012924.03-733236.3 & 0.4 &$\phantom{1}0.26 \pm 0.04$& $ \phantom{-}0.10\pm0.18$ & $ 1.00\pm0.00$ & H,B,S & ? \\
3 &  J012930.17-733310.7 & 0.3 &$15.02 \pm 0.32$& $-0.26\pm0.03$ & $ 0.42\pm0.03$ & B,S,H &QSO\,J012930-733311 \\
4 &  J012930.98-733344.2 & 0.5 &$\phantom{1}0.08 \pm 0.03$& --& --& S,B& cluster center?  \\
5 &  J012931.29-733342.2 & 0.5 &$\phantom{1}0.06 \pm 0.02$& --& --& S &  O9.5V star?\\
6 &  J012935.13-733242.8 & 0.4 &$\phantom{1}0.18 \pm 0.04$& $-0.91\pm0.18$ & $ 0.50\pm0.19$ & S,B & ? \\
7 &  J012935.79-733233.1 & 0.4 &$\phantom{1}0.21 \pm 0.04$& --& $ 0.43\pm0.19$ & S,B & galaxy `G372' \\
8 &  J012939.52-733355.3 & 0.5 &$\phantom{1}0.25 \pm 0.04$& $ \phantom{-}0.08\pm0.20$ & $ 0.75\pm0.19$ & B,S,H& ? \\
9 &  J012942.31-733353.5 & 0.5 &$\phantom{1}0.17 \pm 0.03$& --& --& B,S,H& galaxy
\enddata
\tablecomments{Definition of the {\em Chandra} bands:
0.5--1 (S1), 1--2 (S2), 2--4 (H1), and 4--8~keV  (H2), 
with S\,$=$\,S1\,$+$\,S2, H\,$=$\,H1\,$+$\,H2, and B\,$=$\,S\,$+$\,H.  
Columns: (1) Generic source number;
(2) Adopted \cxo\ identification; 
(3) Positional uncertainty (1$\sigma$) calculated from the maximum likelihood centroiding error and an
approximate off-axis angle ($r$) dependent systematic error: 
$0\farcs2+1\farcs4(r/8^\prime)^2$,
\citep[an approximation to Fig.\,4 from][]{Feigelson2002}, and added in quadrature;
(4) On-axis source broad-band count rate (the sum of the exposure-corrected count rates in the four bands);
(5,\,6) The hardness ratios, defined as 
${\rm HR}=({\rm H-S2})/({\rm H+S2})$, and ${\rm HR1}=({\rm
  S2-S1})/{\rm S}$; 
(7) `B', `S', or `H' indicates the band in which a source is detected
and the source of the quoted position in Col.\,2; (8)
Information on counterparts, see the Appendix for further details;
}
\end{deluxetable*}

In accordance with other results found for the SMC clusters 
\citep[e.g., NGC\,346,][]{Massey1995, Sabbi2008, DeMarchi2011}, 
this suggests that the star-formation process is similar between 
the SMC and  the Galaxy  \citep[e.g.][]{Chabrier2003}.
However, it is not yet clear how low metallicity of the SMC affects 
the physical properties of the stars. 
In massive stars the wind momentum may 
be depend on metallicity \citep[e.g.][]{puls1996}.  
In young low-mass stars the metallicity effects could 
lead to different coronal and accretion properties. 
Additionally, a metal-poor environment
might lead to reduced radiative losses from X-rays, thus affecting disk
formation.  For example, \citet{Yasui2009} suggested that stars
forming in a low-metallicity environment experience disk dispersal a
few Myr earlier compared to those with solar abundances.  The
dynamo process which powers X-ray coronae may also be affected by
metallicity \citep[e.g.][]{Pizzolato2001}.

There are only very few deep X-ray observations of star-forming regions 
in environments which  are very different to our Galactic neighborhood 
\citep[e.g.][]{Car2012}. Previous X-ray observations of star-forming regions in the 
SMC were not sensitive enough to detect the X-ray emission from 
low- and solar-mass stars \citep[e.g.][]{Naze2003}. 
The new \cxo\ data presented
here provide us with an opportunity to investigate the X-ray behaviour
of young stellar populations in the low metallicity of the SMC for the
first time.

\medskip In this paper we present the \cxo\ observations, which
have revealed extended X-ray emission in NGC\,602a. Section~2
presents the observations, data reduction, the detected point-sources
and the details of the extended X-ray emission. The X-ray emission from 
massive stars is considered in Section~3.
Section~4 discusses  
the possible origins of the extended emission, with a summary 
given in Section~5. In Appendix the details about the
X-ray point sources are provided, and the neutral column density towards
\ngc\ is deduced. 

\section{Observations and Data Reduction}\label{sect:observations}

The data were obtained with the \cxo\ ACIS-I detector. The
observations comprised 11 exposures acquired between 2010~March~31
and 2010~April~29 with an effective exposure time of 290.7\,ks. A
combined optical, IR and X-ray image of the cluster is shown in
Fig.\,\ref{fig:combi}.

The \cxo\ data-reduction package {\sc ciao}\,(v4.3) was used for
calibration of the X-ray data and extraction of the spectra\footnote
{All X-ray spectra discussed here were analyzed using the {\sc xpsec}
v12  spectral fitting package \citep{xspec1996}.}. Point-like sources
were detected in three broad bands (`S':~0.5-2.0~keV,
`H':~2.0-8.0~keV, `B':~0.5-8.0~keV) using the procedures from
\citet{Wang2004}.

The detection procedures use a combination of algorithms:
wavelet detection, a `sliding-box' method, and maximum likelihood
centroid fitting. The source count rates are estimated from the
net counts within the 90\% energy-encircled radius determined by the
telescope/ACIS point-spread function \citep[PSF,][]{PSF2000}. The PSFs
are different in the soft and hard bands, therefore the
point-source detection was undertaken in the individual bands. The
detection limit is characterized by a threshold probability (we used
$1\times 10^{-6}$), which describes the probability of a false
detection of a random variation above the background as a point
source.

For our observations we used the position-dependent 70\%\
energy-encircled radius as the detection aperture \citep{PSF2000},
corresponding to $\sim$1\farcs2 (0.3\,pc) in NGC\,602a. The adopted
threshold probability allowed us to detect point sources with a
count-rate of 0.05\,cts\,ks$^{-1}$ in a broad band, corresponding to
14 source counts (3$\sigma$ detection). No flaring activity within
\ngc\ was detected during our observations.

\subsection{Point sources detected in NGC\,602a}
\label{pointsources}
Nine point sources were detected within 2$'$ ($\sim$\,32\,pc) of the
core of \ngc. These are summarised in Table~\ref{tab:ngc602src} and
their locations are overlaid on an optical {\em HST} image in
Fig.\,\ref{fig:hstcon}. Source 2 has no optical counterparts
within a radius of $1''$ \citep[cf. the source catalogue
from][]{Schmalzl2008}; we believe  it to be a background source.  
The counterparts of the remaining sources are discussed in more detail 
in the Appendix.  Two (sources~4 and 5) are located in the dense core 
of NGC\,602a, with source~5 apparently coincident with a late O-type star.  
Source~3 is a newly-discovered quasar, QSO\,J012930-733311. Using our own
optical spectra obtained with VLT-FLAMES we determined its 
redshift as $z$\,$=$\,2.4. 

We use the X-ray spectral analysis of QSO\,J012930-733311 
to estimate the neutral hydrogen column density towards \ngc\
(see Section\,\ref{sec:qso}). Throughout this paper we use
a two component absorption model, with Galactic 
foreground absorption at 6\,$\times$\,10$^{20}$\,cm\,$^{-2}$
\citep{Wilms2000} and a second component with SMC abundances at 
2\,$\times$\,10$^{21}$\,cm$^{-2}$.

\subsection{Extended X-ray emission}

\begin{figure*}
\centering
\includegraphics[width=1.5\columnwidth]{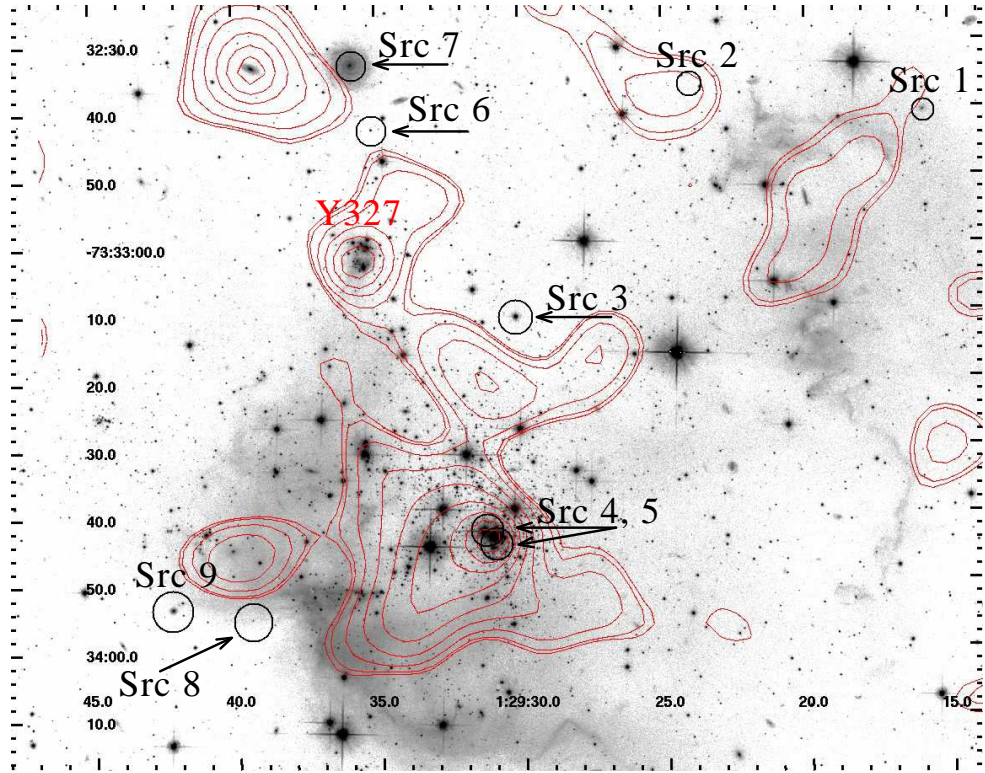}
\caption{{\em HST} image of \ngc\ (F658N filter, north is at the top,
  east to the left). Black circles are point sources detected by \cxo\
  (Table\,\ref{tab:ngc602src}). Red contours show areas of extended
  X-ray emission at the 3\,$\sigma$ level.  
  From north-to-south, the three areas of notable extended
  X-ray emission correspond to a background spiral galaxy and
  sub-clusters\,2 and 1 from \citet{Goul2012}. North is up, and east is to the left.}
\label{fig:hstcon} 
\end{figure*}

\begin{figure*}
\centering
\includegraphics[width=1.9\columnwidth]{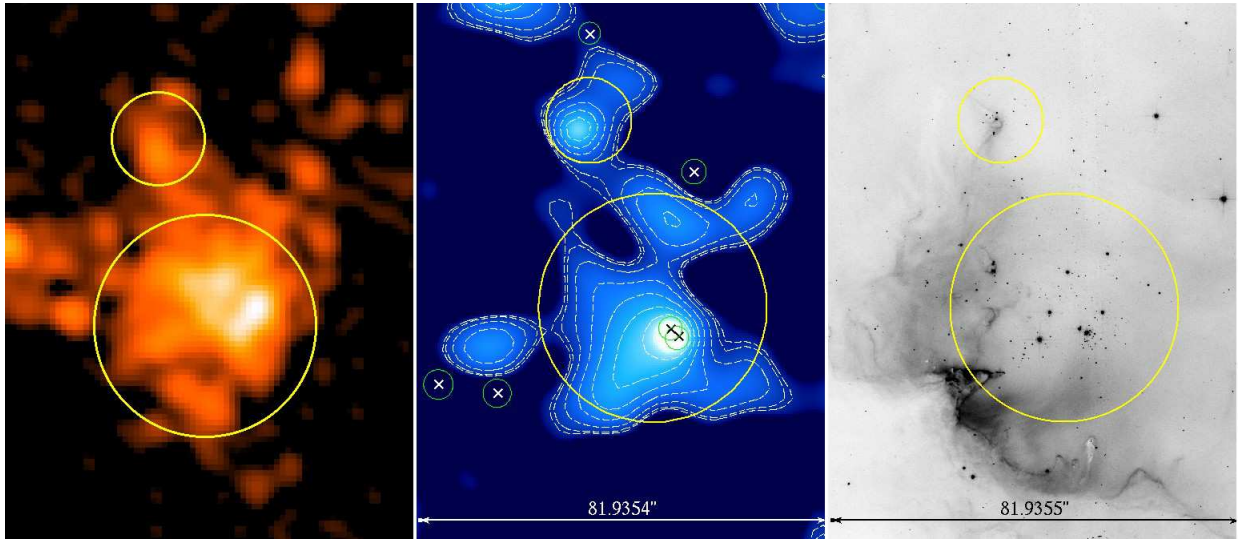}
\caption{{\em Left panel:} Stellar density map of the low-mass PMS
  stars in the two most dense sub-clusters in \ngc\ (image courtesy of
  D.\,A.\,Gouliermis, {\changed see 
Figs.\,5 \& 9 in \citet{Goul2012} for details}). 
{\em Middle panel:} Adaptively-smoothed \cxo\
  image of a part of \ngc. White contours trace the extended X-ray
  emission on a square-root scale, from
  5\,$\times$\,10$^{-4}$\,cnt\,s$^{-1}$\,arcmin$^{-2}$ to
  8\,$\times$\,10$^{-4}$\,cnt\,s$^{-1}$\,arcmin$^{-2}$.  The encircled
  crosses show point sources detected by \cxo, but note that all but
  these two point sources have been subtracted prior from this image.
  {\em Right panel}: {\em HST} F658N image
  of the same area for comparison. Sub-clusters\,1 and 2 are shown in
  all three panels by the overlaid yellow circles, with centers and
  radii from Table\,1 of \citet{Goul2012}. }
\label{fig:dens} 
\end{figure*}

The high angular resolution of \cxo\ allowed us to disentangle bright
point sources from areas of extended X-ray emission. After subtraction
of the point sources, inspection of the adaptively-smoothed X-ray
images reveals apparent areas of extended, diffuse emission in \ngc\
(see Figs.\,\ref{fig:combi}\,--\,\ref{fig:dens}). 
The largest area of extended emission is centered on
the central part of \ngc, with a more compact area of emission
coincident with the sub-cluster of YSOs Y327 \citep{Carlson2011}.  
The third area is to the north of the cluster
and coincides with a galaxy seen in the {\em HST} images (see
Fig.\,\ref{fig:hstcon}).

Formal detection of extended X-ray emission must satisfy two criteria:
(1) the extended emission shall be above the background fluctuations
and, (2) it should be truly extended, with a radial profile that is
wider than that of the PSF.  The first criterion is checked using image
contours at the 5\,$\sigma$ and 3\,$\sigma$ levels (e.g.,
Fig.\,\ref{fig:hstcon}). {\changed Once an area is identified as possible
extended emission, its extent is checked by building a radial profile,
which is then compared with the profile from a point-like source of 
similar brightness (see Fig.\,\ref{fig:radprof}).}
\citet{Goul2012} showed that \ngc\ is hierarchically structured, and
identified sub-clusters that differ in age, mass, and size (their
Table~1).  Two of the areas of extended X-ray emission coincide with
their two most populous sub-clusters (see Fig.\,\ref{fig:dens}), which
are now discussed in more detail.

\subsubsection{Sub-cluster\,1: the core of NGC\,602a}\label{sc1}

\begin{figure}
\centering
\includegraphics[width=0.99\columnwidth]{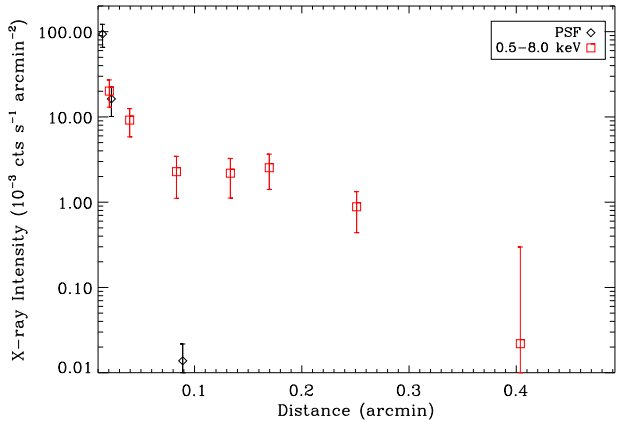}
\caption{Radial profile of the extended X-ray emission in
  sub-cluster\,1 of \ngc. The center is at
  $\alpha$\,$=$\,01\fh29\fm31\fs104,
  $\delta$\,$=$\,$-$73$^\circ$33$'$44\farcs19. The data points (red
  squares) represent bins with a signal-to-noise ratio $\geq 2.5$,
  with the exception of the last point. For comparison, the PSF of a
  point source (Source~4 from Table~1) in \ngc\ is shown (black
  squares).}\label{fig:radprof} 
\end{figure}

At least seven OB-type stars and numerous lower-mass stars reside in
`sub-cluster\,1' \citep{Goul2012}, with the stellar density
distribution comparable to the distribution of extended X-ray emission
(see Fig.\,\ref{fig:dens}). A comparison of the radial profile of the
X-ray emission filling the interior of sub-cluster\,1 with the ACIS
PSF is shown in Fig.\,\ref{fig:radprof}, demonstrating that the
X-ray emission is indeed extended.

After removal of the detected point sources, an X-ray spectrum was
extracted from an area of $\approx$18\,pc$^2$ (as shown by the solid
circle in Fig.\,\ref{fig:extend}). 
{\changed We investigated in detail the effects of the background. The spectra 
obtained using different background areas are similar, and they are 
always above the background level. E.g.\ the total (together with 
background) count rates (over 0.5-8.0 keV band)  of extended X-ray 
emission from sub-cluster\,1 is $(5.0\pm 1.0)\times 10^{-4}$\,counts\,s$^{-1}$, 
while the ambient background count-rate in the same band is 
$(0.6\pm 1.0)\times 10^{-4}$\,counts\,s$^{-1}$. Therefore, the source 
spectrum is about 3$\sigma$ above the background spectrum.}

\begin{figure*}
\centering
\includegraphics[width=1.5\columnwidth, angle=-90]{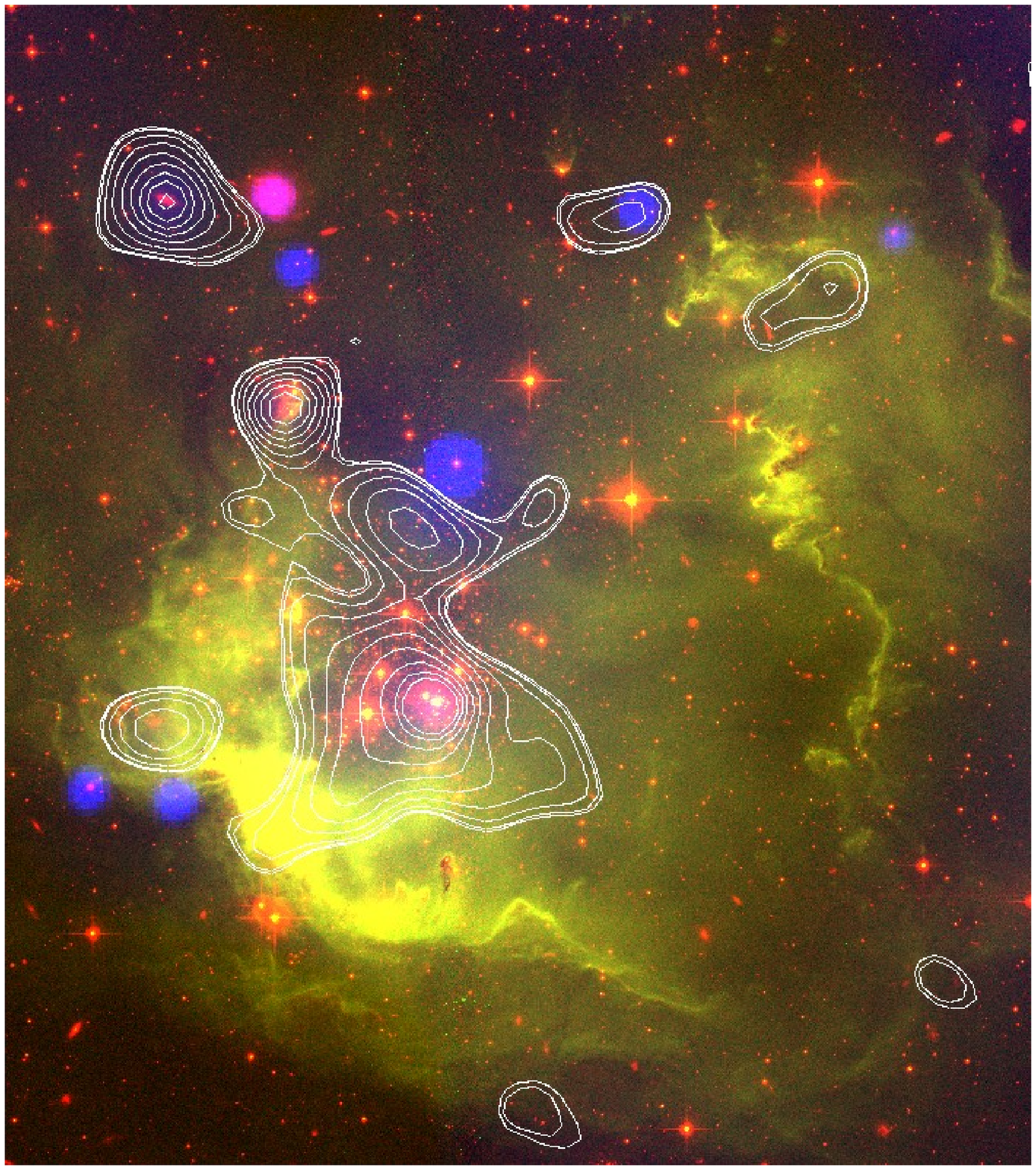}
\caption{Color-composite image of \ngc. {\em HST} images are in red
  (F814W filter) and green (F658N filter). The blue image is the
  \cxo\ 0.5-8.0\,keV band with X-ray point sources removed and
  adaptively smoothed.  The contours tracing the extended X-ray
  emission are the same as in the middle panel in
  Fig.\,\ref{fig:dens}. Image size is $\approx 2.6'\times 2.6'$ 
($\approx 42$\,pc$\times$42\,pc). North is up, and east is to the left.}
\label{fig:2hstcon} 
\end{figure*}

{\changed To derive the spectral parameters, we first attempted a spectral 
fit with a two-temperature plasma model. Unfortunately, due to the small 
number of the spectral counts, the 2-T model parameters could not constrained. 
Therefore, we use an {\sc apec} isothermal plasma model, adopting SMC 
abundances and the absorption model as obtained from the fits to 
the nearby, newly-discovered QSO\,J012930-733311 (see Section~\ref{sec:qso}). 
The best fit one-temperature model has kT\,$=$\,2.1\,$\pm$\,1.3\,keV with an
emission measure of (10\,$\pm$\,3)\,$\times$\,10$^{55}$\,cm$^{-3}$, 
The model observed flux is 
2\,$\times$\,10$^{-15}$\,erg\,s$^{-1}$\,cm$^{-2}$ in the 0.5-8.0\,keV 
and corresponding  luminosity of the extended X-ray 
emission in sub-cluster\,1 is 
$L_{\rm X}\,\approx\,1 \times10 ^{33} - 2 \times 10 ^{33} $\,erg\,s$^{-1}$.}

\subsubsection{Sub-cluster\,2: the tip of ``elephant trunk''}\label{sc2}

Extended X-ray emission (over an area of $\sim$\,12\,pc$^2$ or 
$\sim$\,55\,arcsec$^2$) is also detected from `sub-cluster\,2', 
the second most populous sub-cluster in the region \citep{Goul2012}. 
Located at the tip of an `elephant trunk' associated with star formation 
and on the northeastern limb of the N\,90 H\,{\sc ii} region, 
sub-cluster\,2 is a good example of a compact cluster at an earlier 
evolutionary stage than NGC\,602a.

Two IR-bright, massive YSOs were identified in 3.6\,$\mu$m {\em
  Spitzer} images by \citet[][their sources 54 and 57]{Goul2007}. 
\citet{Carlson2011} collectively refer to these, and other nearby 
bright sources, as `Y327' (this label is also used to identify the 
sub-cluster in Fig.\,\ref{fig:hstcon}) and identified them
as Class~I YSOs (younger than $\sim$10$^{5}$\,yr). Class~I YSOs 
are surrounded by  in-falling dusty envelopes, which continue 
to accrete mass and produce outflows \citep{Adams1987}.


Only 22 counts were detected from the \cxo\ observations of
sub-cluster\,2, none of which were at energies below 1\,keV,
confirming that this object is highly embedded. Comparing
the number of counts at different radii with those expected from a
point source confirms that X-ray emission is extended 
at this position. It is difficult to estimate the intrinsic
luminosity as we do not know the column density and extinction law in
this region. Adopting an {\em ad hoc} $N_{\rm
  H}$\,$=$\,10$^{22}$\,cm$^{-2}$ with a thermal plasma with
temperature $\sim$2\,keV we estimate the {\it model dependent} X-ray
luminosity of sub-cluster\,2 to be {\changed $L_{\rm
  X}$\,$\approx\,$6\,$\times$\,10$^{32}$\,erg\,s$^{-1}$ in the
0.5-8.0\,keV band}.

\section{X-ray Emission from Massive Stars in \ngc}
\label{sec:mas}

On the basis of \emph{HST} photometry, \citet{Schmalzl2008} identified
the 10 brightest stars of \ngc\ as massive stars (i.e.\ with inferred
spectral types of B0.5 or earlier). Among all these stars, only the
O9.5V star SGD\,13 is marginally detected in our observations 
(see discussion in Section\,\ref{sec:sgd13}). 

For Galactic single and binary OB stars the bolometric and X-ray
luminosities are  seen to correlate as $\log{L_{\rm X}/L_{\rm
bol}}$\,$\sim$\,$-$7 \citep[e.g.][]{Seward1979,
Pallavicini1981,Berg1997,Oskinova2005,naze2009}.  X-ray detections of
normal massive stars in other galaxies are scarce so it is unknown how
this correlation scales with environment. Given that the strength of
stellar winds decreases with metallicity
\citep[e.g.][]{puls1996,Vink2001,mokiem2007,Graf2008}, we expect that
O-type stars in the SMC are less X-ray luminous than in the Galaxy.

To estimate an upper limit to the X-ray luminosity of undetected
OB-type stars in \ngc, we adopt the line-of-sight neutral hydrogen
value from analysis of the nearby QSO X-ray spectrum
(Section~\ref{sec:qso}), and assume that the X-ray spectral model is
similar to that found typically in Galactic massive stars, i.e., an
optically-thin plasma in collisional equilibrium, characterized by
$T$\,$\sim$\,5\,MK.  With these assumptions, any massive stars not
detected by \cxo\ in \ngc\ should have X-ray luminosities of
$\lsim$\,1.5\,$\times$\,10$^{32}$\,erg\,s$^{-1}$. Adopting this limit,
if massive stars in the SMC were as X-ray luminous as their Galactic
counterparts, then only the most massive stars (with spectral types
earlier than O4~V) would have been detected.

Using optical spectroscopy with VLT-FLAMES of massive stars in and around
NGC\,602a (Hainich et al.  in prep.) we found that Sk\,183, 
is the only star in NGC\,602a with such an early spectral type.  
However, it was not detected in our X-ray observations.  Given the 
detection limit calculated above, the maximum X-ray luminosity for 
Sk\,183 (compared to its bolometric luminosity) is 
$\log{L_{\rm X}/L_{\rm bol}}$\,$\lsim$\,$-$7.2.

After Sk\,183, the next most massive stars known in the cluster are
late O-type dwarfs \citep[e.g.][Hainich et al. in
prep.]{Hutchings1991}.  The X-ray luminosities of Galactic late O-type
stars are typically much lower than the detection limit of the \cxo\
observations in NGC\,602a. For instance, both $\zeta$\,Oph (O9~V) and
$\mu$\,Col (O9.5~V) have $L_{\rm
  X}$\,$\approx$\,($1.2-1.3$)\,$\times$\,10$^{31}$\,erg\,s$^{-1}$
\citep{Oskinova2001, Huen2012}; in both cases this corresponds to
$\sim$10$^{-7}$ of their bolometric luminosity.

The correlation of X-ray to bolometric luminosity can have a spread of
up to 1\,dex \citep[e.g.][]{Naze2011}.  However, if $L_{\rm X}$
exceeds $10^{-7}L_{\rm bol}$ it strongly indicates that the star is
either a colliding-wind binary 
\citep[e.g.][]{Stevens1992, Antokhin2004, 
Pollock2006}, or has a magnetically-confined wind 
\citep[e.g.][]{BabMon1997, Donati2002}\footnote{We note
that binary stars and magnetic stars can also exhibit 
average X-ray luminosities and soft X-ray spectra
\citep{Drake1994, Oskinova2005, naze2009, Oskinova2011}}.
In these cases one would also expect relatively hard X-ray emission
\citep[e.g.][]{Gagne2005, Ignace2012, naze2012}.
Even more significant X-ray fluxes would be expected from massive
stars with a degenerate (neutron star or black hole) companion, for
which $\log{L_{\rm X}/L_{\rm bol}}$\,$\lsim$\,$-$4\,...\,$-$3
\citep{Liu2005}.

The fraction of binaries among O-type stars is very high (up to 70\%), 
especially in young  clusters \citep{chini2012,sana2012}. It is likely,
that there are binaries among massive stars in \ngc. The non-detection 
of these stars in X-rays  shows that a colliding wind binary phenomenon
is quite rare among dwarf O-type stars. This may be due to the weakness
of their stellar winds, as X-ray luminosity scales with $\dot{M}^2$
\citep{Stevens1992}. Therefore, the upper limits on X-ray emission from 
massive stars in \ngc\ are in agreement with expectations. 

Given the deep optical and IR imaging available, the census of massive
stars in NGC\,602 can be assumed to be complete (to first order) and,
aside from the notable high X-ray luminosity of source~5 ($L_{\rm
  X}$\,$=$\,3$\times$\,10$^{32}$\,erg\,s$^{-1}$, see discussion in
Section\,\ref{sec:sgd13}), it seems extremely unlikely that the extended
emission in sub-cluster\,1 could be accounted for by an
unresolved/undetected population of high-mass stars.

\section{Origins of the extended emission}
In this section we discuss the origins of the extended X-ray emission 
detected in sub-clusters\,1 and 2.

\subsection{Extended Emission from a Wind-blown Bubble?}\label{wbb}

The intracluster gas in a stellar cluster can be heated by the
mechanical energy input from stellar winds. This can result in an
outflow of a hot cluster wind and creation of a wind-blown bubble
(WBB). Indeed, the diffuse X-ray emission from hot gas is often
observed in star-forming regions \citep[e.g.][]{Townsley2003,Gudel2008,Townsley2011}.
The density and the temperature of the hot, X-ray emitting gas is
largely determined by the kinetic energy input and depends
sensitively on the cluster mass, age, and metallicity
\citep{Chu1990, Stevens2003, Oskinova2005}. Theoretical models from
\citet{Silich2005} indicate that the X-ray production efficiency by
the diffuse intracluster gas is generally low, with $L_{\rm X}/L_{\rm
  mech}$\,$\ll$\,1\%, as found by \citet{Smith2005}. In the absence of
recent SNe, the heating of the WBB is entirely due to the stellar 
winds, and is determined by the total number of massive stars and 
kinetic energy of their winds.  

As was already discussed above, ten massive stars in \ngc\ are known -- 
an average number for a cluster of such mass. The most massive star in
\ngc, Sk\,183, is $\sim$45$''$ ($\sim 12$\,pc) to the 
northwest of sub-cluster\,1. 
The earliest type stars known in sub-cluster\,1 are three late O-type
dwarfs (see Sections\,\ref{sec:sgd12}, \ref{sec:sgd13}). 

To estimate the combined feedback  from stellar 
winds in this sub-cluster we need to know their mass-loss rates and wind
velocities.  \citet{Mokiem2006} investigated wind parameters for
late O-type stars in the SMC.  Their mass-loss rates (in the range of 
10$^{-7}$ to 10$^{-8}$\,$M_\odot$\,yr$^{-1}$) were obtained from analysis 
of the observed H$\alpha$ profiles. In late O-dwarfs, H$\alpha$ is in
absorption and does not significantly depart from a pure photospheric 
profile. Combining optical and UV spectroscopy helps to set more
stringent limits on mass-loss rates. The UV spectra obtained with 
{\em HST} STIS combined with optical spectroscopy were used by 
\citet{bouret2003} to analyse O-dwarfs in the SMC.  They derived
mass-loss rate $1\times 10^{-10}\,M_\odot$\,yr$^{-1}$ and 
$v_\infty\sim 1000$\,km\,s$^{-1}$ for a O9.5V-type star. Similarly, 
combining optical and UV spectroscopy, \citet{Martins2004} determined 
that the winds of O-dwarfs in the SMC cluster N\,81 (located in the 
SMC Wing) are weak, with mass-loss rates typically lower than  
${\rm few} \times 10^{-9}\,M_\odot\,{\rm yr}^{-1}$ and rather low 
wind velocities. \citet{Oskinova2007} showed that to derive accurate 
mass-loss rate from UV resonance lines, stellar wind clumping has to
be correctly accounted for in the radiative transfer and that in this
case, the consistent mass-loss rates are obtained from the analysis of 
H$\alpha$ and the UV lines. These mass-loss rates are only factors 2-3
lower than the theoretical predictions \citep[e.g.][]{Mokiem2006}.  
The new Monte-Carlo fully 3-D stellar wind models that account for 
the interclump medium confirm this result \citep{surlan2012}.  
 
With all these in mind, we adopt 
$\dot{M} = 10^{-8}\,-\,10^{-9}\,M_\odot$\,yr$^{-1}$ 
and $v_\infty = 1000$\,km\,s$^{-1}$ for the three O9.5V stars in
sub-cluster\,1.  
The other stars in NGC\,602a are early B-type (or later) dwarfs, so 
their wind kinetic energies will be even smaller \citep[see wind 
analysis of B-dwarfs in][]{Oskinova2011}. Thus, the combined input of
kinetic energy from stellar winds in sub-cluster\,1 is 
$L_{\rm mech} \lsim 10^{33}- 10^{34}$\,erg\,s$^{-1}$, which is too small to
power the detected extended X-ray emission. 
This conclusion is supported by the morphology of the extended X-ray
emission.  Typically, WBBs have limb-brightened morphologies, where
the diffuse X-ray emission is encompassed by, e.g., H$\alpha$ shells
\citep{Chu1990,Toala2012}, but no such shell is seen around sub-cluster\,1
(e.g., Fig.\,\ref{fig:hstcon}).

\begin{figure}
\centering
\includegraphics[scale=0.6]{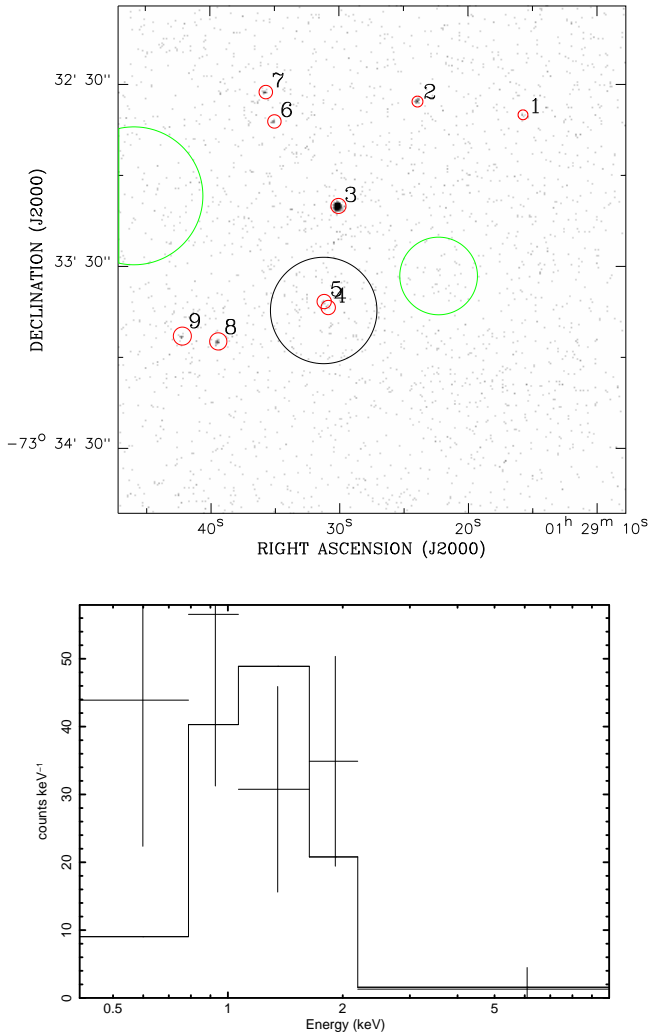} \\
\includegraphics[scale=0.33,angle=-90]{diff1T.ps}
\caption{{\em Upper panel:} ACIS-I raw event map with the spectral
  extraction regions shown. Red circles: X-ray point sources; black
  circle: the region extracted to investigate the extended emission
  associated with sub-cluster\,1 from \citet{Goul2012}; green circles:
  the regions used to extract background spectra. {\em Lower panel:}
  The spectrum of extended emission from sub-cluster\,1 and its best
  fit isothermal ({\sc apec}) plasma model (see Section~\ref{sc1} for the 
model parameters). The spectrum is grouped such that the signal-to-noise
  ratio S/N\,$=$\,2 in each energy bin. The best-fit model is found using
  Cash-statistics based fitting.}
\label{fig:extend} 
\end{figure}

{\changed Beyond these considerations, we are unable to firmly exclude a
contribution to the extended emission in sub-cluster\,1 from a WBB.
This is because, as noted in Section~\ref{sc1}, the quality of the X-ray
spectrum does not allow us to test multi-temperature spectral 
fits  (Fig.\,\ref{fig:extend}). Also, we
note that the experience from X-ray observations in the ONC suggests
that \cxo\ is not ideally suited for detection of WBBs because of 
its rather low sensitivity for soft X-rays.}  
The diffuse, hot ($T\sim 2$\,MK) plasma discovered in the ONC by
\citet{Gudel2008} was from {\em XMM-Newton} observations, and was not
detected by \cxo\ \citep{Feigelson2005}. In part this was due to lower
sensitivity of \cxo\ to soft X-rays.  Moreover, the ONC is absorbed by
a `veil' of neutral gas \citep{Odell2011} which was evident in the
$N_{\rm H}$ measurements from \cxo; in the region further out where
the diffuse emission was observed by {\em XMM-Newton}, the H~{\sc i}
column density is much smaller.

Lastly, we also note that \citet{Gudel2008} explained all of the hard,
extended X-ray emission observed in the ONC as being due to PMS stars.
\citet{Arthur2012} has recently presented hydrodynamic models
for the expansion of WBBs inside evolving H\,{\sc ii} regions,
adopting stellar wind parameters and a rate of ionizing photons
appropriate to $\theta^{1}$\,Ori\,C. She found that the model
over-predicts the emission at higher energies compared to the
observations. New physical ingredients (e.g., mass-loading,
non-spherical models) may be required to reconcile WBB models for
Orion with the observations.  Thus, to firmly test a WBB model for
\ngc\ we require dedicated theoretical models to arrive at robust
predictions about their X-ray emission. 

\subsection{Sub-cluster\,1: X-ray Emission 
from pre-Main Sequence Stars}\label{pms}

X-ray observations of Galactic star-forming regions have revealed that
{\em essentially all}~low- and solar-mass PMS stars are X-ray sources. For
example, \cxo\ observations of the ONC detected 98.5\%\ of the PMS stars
known from optical and IR studies \citep{Preibisch2005}. Thus,
sufficiently deep X-ray observations of star clusters should be
sensitive to the PMS population and the numerous (but individually
faint) sources spread over a cluster may dominate the extended X-ray
emission. This is especially true in a young clusters before the 
first supernovae and when the input  of kinetic energy from massive stars 
is low due to their relative youth 
\citep[e.g.][]{Oskinova2005} -- sub-cluster\,1 in \ngc\ is
at such evolutionary stage.

There is a rich population of PMS stars in \ngc\ \citep{Carlson2007,Schmalzl2008}. 
\citet{Carlson2011} commented that low-mass PMS stars (with
0.6\,$<$\,$M$\,$<$\,3\,$M_\odot$) were the most remarkable feature
in the optical color-magnitude diagram of \ngc, with ages of generally
less than $\sim$5\,Myr.  While some clumps of the PMS stars appear in
the dusty outskirts of the cluster, the majority are concentrated in 
sub-cluster\,1 \citep[with an excess of 1\,000 observed stars,][]{Goul2012}.

We still know little about activity and accretion onto lower mass PMS
stars in a low metallicty environment such as the SMC with except
that their H$\alpha$ emission properties appear to be relatively normal
\citep{DeMarchi2011}. We hypothesize that the
properties of PMS stars in \ngc\ are similar to those in the
ONC.  To test this hypothesis, we check whether we can explain 
the observed extended X-ray emission by {\em assuming} that stars 
of similar mass and age have the same X-ray properties in the ONC 
and \ngc.

The \cxo\ observations of the ONC allowed \citet{Preibisch2005} to
obtain X-ray luminosities of stars in different mass bins. They found
that for PMS stars with ages in the range $\sim 0.1 - 10$\,Myr stellar
activity decays only mildly with age. We employ their statistical
correlations (from their Table~1) between stellar mass, age, and X-ray
luminosity in the 0.5-8\,keV band and in four mass bins, namely:
$0.1-0.2$, $0.2-0.4$, $0.4-1$, and $1-2\,M_\odot$.

For simplicity, we assume that all PMS stars in \ngc\ are coeval with ages 
of 4\,Myr. 
At this age, intermediate mass  stars ($2-8\,M_\odot$), would
normally be  recognized as a Herbig-type objects. Because HAeBe stars are not
expected  to possess an outer convective layer, X-ray emission detected
from these objects  is often attributed to low-mass companions.  There
is however evidence to suggest that  in at least  some HAeBe stars 
intrinsic X-ray emission can originate from solar-like magnetic coronae
or  magnetically confined winds or shock heated plasma in the jet or
wind  \citep{Telleschi2007,Gunther2009}. Whatever  the origin of X-ray
emission from  intermediate mass young stars is, their X-ray luminosity is
moderate, not exceeding  a few$\times 10^{29}$\,erg\,s$^{-1}$.
Therefore, we assume that these stars are X-ray dim,  and that their
X-ray luminosity can be attributed to the less massive  companion in
binaries \citep[e.g.][]{EvansN2011}.

We assume a standard broken power law form for the IMF \citep{Kroupa2001} and
randomly sample stars with masses between $0.1\,M_\odot$ and $50\,M_\odot$ 
(similar to the mass range for which \citet{Carlson2011} estimated the mass 
of the cluster) from this IMF. The results are not very sensitive to the 
high-mass cutoff used due to the small number of stars in these bins
For a  cluster mass of $\approx 2250\,M_\odot$, 
the number of stars between $0.1\,M_\odot$ and $2\,M_\odot$ is 
$\approx 3400$. Their cumulative X-ray luminosity is 
$L_{\rm X}\approx 2.3 \times 10^{33}$\,erg\,s$^{-1}$. 

To check how sensitive  the X-ray luminosity is to the cluster mass, we also  
considered a lower limit of $\approx 1600\,M_\odot$ for the total cluster 
mass as found by \citet{Cignoni2009} from optical data alone. In this case, 
the number of stars between  $0.1\,M_\odot$ and $2\,M_\odot$ is $\approx 2500$, 
and the total X-ray luminosity is  $L_{\rm X}\approx 1.7 \times 10^{33}$\,erg\,s$^{-1}$. 

The cluster mass reported by \citet{Carlson2011} and \citet{Cignoni2009} 
is for the whole of \ngc. Based on the results of \citet{Goul2012}  we might
expect the mass of sub-cluster 1, and hence the number of stars in this
region, to be $\approx 25$\%\ smaller, so the above numbers would need to
be scaled down accordingly.

With these assumptions, the model X-ray luminosity of PMS stars in \ngc\ 
is $L_{\rm X}$\,$\lsim$\,2\,$\times$\,10$^{33}$\,erg\,s$^{-1}$
(0.5-8\,keV band), in excellent agreement with the result from the
spectral fits of the extended emission from sub-cluster\,1 
(see Section\,\ref{sc1}).
 
The spectral shape of extended X-ray emission from sub-cluster\,1  
also agrees well with the global X-ray properties of the low-mass population 
associated with the ONC from \citet{Feigelson2005}. They found that 
a composite spectrum of low-mass stars in the ONC can be well represented 
by a two-temperature spectral model with $kT_1$\,$=$\,0.5\,keV and
$kT_2$\,$=$\,3.3\,keV (with non-solar abundances), in modest
agreement with those from the spectral fits in Section~\ref{sc1}.

Thus, we believe that the extended X-ray emission is explained as
originating from the unresolved population of low-mass, X-ray active
PMS stars.  

\subsection{Sub-cluster\,2: \\ 
X-ray emission from embedded class I YSOs}

The extended X-ray emission is also detected from the second most
populous sub-cluster (see Section \ref{sc2}). We attribute this emission
to the unresolved population of Class\,I YSOs.

Despite the large column densities in Galactic star-forming regions,
Class\,I YSOs are routinely detected in X-rays, with typical
luminosities of ${L_{\rm X}}$\,$\lsim$\,10$^{30}$\,erg\,s$^{-1}$
\citep[e.g.][]{Getman2002,Preibisch2005} and $L_{\rm
  X}$\,$\approx$\,10$^{-4}$\,$L_{\rm bol}$
\citep[e.g.][]{Winston2010}.  Magnetic effects as well as accretion
and jets are the likely key mechanisms responsible for X-ray
production \citep{For2007}.

The most energetic YSO at the tip of the ``Pillar of Creation'' in the
Eagle Nebula has $L_{\rm
  X}$\,$\approx$\,3$\times$\,10$^{31}$\,erg\,s$^{-1}$
\citep[e.g.][]{Guarcello2012}, so dozens of such energetic YSOs would
be needed to explain the X-ray luminosity of sub-cluster\,2.  Given
that sub-cluster\,2 is embedded in the dusty ridge, the absorption of
X-rays will depend on the abundances and dust properties of that
material \citep{Wilms2000,Gordon2003}, thus the estimates of the
luminosity may be affected by the adopted extinction.  An alternative
explanation may involve true diffuse X-ray emission, e.g.\ as was suggested
by \citet{Oskinova2010} who reported hard, diffuse X-rays from the Galactic
star-forming region ON\,2.  Nevertheless, despite the large
uncertainty in the estimated X-ray luminosity, one of the key results
from the \cxo\ observations is the first X-ray detection of an
embedded cluster of YSOs in the SMC.

\section{Discussion}


X-ray observations of star-forming regions in the Galaxy have provided us with a
high-energy perspective on stellar nurseries via studies of both individual
stars and their overall stellar populations.  The \cxo\ observations
in the Wing of the SMC presented here allowed to obtain a
high-energy view of a star-forming region in a very different environment.  
In addition to detections of point sources, the new observations have
revealed extended X-ray emission emanating from the two most populous
sub-clusters in NGC\,602a.

Optical and IR observations of \ngc\ have shown that the basic
properties of its low-mass stellar population do not appear
significantly different to those in Galactic clusters with comparable
mass and age. The continued star formation in the broken ring around
\ngc\ is also analogous to regions seen in the Galaxy, which are often
considered as an indication of a triggered second generations of star
formation \citep[e.g.][]{Koenig2012}.

Observations of the gas in  the N\,90 H\,{\sc ii}
region demonstrated it to be nearly quiescent \citep{Nigra2008},
suggesting that there has not yet been a supernova explosion in the
cluster\footnote{The non-thermal radio source in the vicinity of 
\ngc\ discussed by Nigra \etal\ as a possible supernova remnant 
is now identified with QSO~J012930-733311}. A supernova 
remnant recently discovered at $\sim$110\,pc to the west of 
\ngc\ \citep{HB2012} is most likely not related to this cluster, but
rather to the general massive star population in the SMC's Wing.
  Stellar winds can therefore be expected to be the principal
source of kinetic energy input into the intracluster medium, but at a
reduced level compared to Galactic stars due to the weaker winds at
the low metallicity of the SMC.  In this context it's notable that the
most massive star (Sk\,183) is not detected in X-rays, in keeping with
its low mass-loss rate \citep{Evans2012}. As might be expected for the
low stellar wind efficiency calculated in Section~\ref{wbb}, we do not
see compelling evidence for a hot, diffuse gas filling the full
cluster volume and originating from a cluster wind or a WBB.  In this
scenario, photo-ionization (which is dominated by Sk\,183) is the
primary driving mechanism in the evolution of the H\,{\sc ii} region
N\,90 \citep[as discussed by][from different arguments]{Carlson2011}.

We therefore suggest that the extended X-ray emission arises from the
unresolved population of low-mass PMS stars (Section~\ref{pms}).  As
no individual PMS star (nor any flaring) is detected, this conclusion is
based on the assumption that the coronal properties of the PMS stars in 
\ngc\ are comparable to the well-studied PMS stars in the ONC.
Adopting the X-ray luminosities from PMS stars in the ONC we calculated
an expected X-ray luminosity for sub-cluster\,1 (on the basis of the known
population of PMS stars). The predicted luminosity matches the
luminosity of the extended emission obtained from our observations.

X-ray emission from PMS stars uniquely traces their {\em magnetic}
activity, and provides an empirical foundation to the theory of
magnetic dynamos -- one of the key ingredient in stellar physics. We
speculate that if the X-ray properties of PMS stars are indeed
comparable in different environments, then other related properties,
such as the formation and evolution of protoplanetary disks, are also
likely to be similar.

The detection of hard X-ray emission from the YSOs residing at the tip
of the `elephant trunk' (sub-cluster\,2) is also significant. Only 22 
counts were detected (with no spectral or temporal information) but, in
combination with the insights obtained from IR studies, such a detection
suggests that the X-ray behaviour for the accretion and magnetospheric
interactions in YSOs in the SMC arise from a similar mechanism to that 
in the Galaxy.






\acknowledgments This study is based on observations obtained by the
\cxo\ science mission, and spectroscopy obtained from ESO program
086.D-0167. This study used software provided by the \cxo\ CXC, and
made use of the NASA Astrophysics Data System Service and the SIMBAD
database. We thank D.\,Gouliermis for providing us with the maps of
stellar density in \ngc. The authors thank Jacco Vink, Leisa K.~Townsley, 
and Helge Todt for useful discussions. We also thank the referee for useful 
and detailed comments which led to considerable improvement of the 
manuscript. Support is acknowledged as follows -- LMO: DLR grant 
50 OR 1101; VHB: SUPA and
NSERC; YN: FNRS, CFWB, ARC, and PRODEX; CJE: Caledonian 80$/$-; MAG:
MICINN grant AYA2011-29754-C03-02, which includes FEDER funds; WS:
DAAD grant A/10/95420; WS and YC: NSFC grant 11233001 and 973 Program
grant 2009CB824800; SS: Conacyt research grant 131913; YHC, JSG, and
RAG: NASA grants SAO GO0-11025X and NNX11AH96G. JSG also thanks 
donors to the University of Wisconsin-Madison College of Letters 
\& Science for partial support of this research.

\appendix
\section{Counterparts to detected X-ray point sources in NGC\,602a}\label{sec:point}

The positions of the nine point sources detected in the \cxo\
observations are overlaid on the optical {\em HST} image shown in
Fig.\,\ref{fig:hstcon}.  As noted in Section~\ref{pointsources},
source~2 lacks a convincing optical counterpart,
we now discuss the other eight in turn.

\subsection{Sources 1 and 8}
Sources\,1 and 8 have faint optical
counterparts. Source 1 is $1\farcs6$ away from SGD\,3373 
\citep[25.6\,mag. in F555W filter;][]{Schmalzl2008}. Source 8 coincides 
with SGD\,3012 \citep[26.8\,mag. in F555W filter;][]{Schmalzl2008}.
We assume that these objects are background AGNs.

\subsection{Source\,3 QSO\,J012930-733311 and
the neutral hydrogen absorption estimates}\label{sec:qso}
\begin{figure}
\centering
\includegraphics[width=0.9\columnwidth, angle=-90]{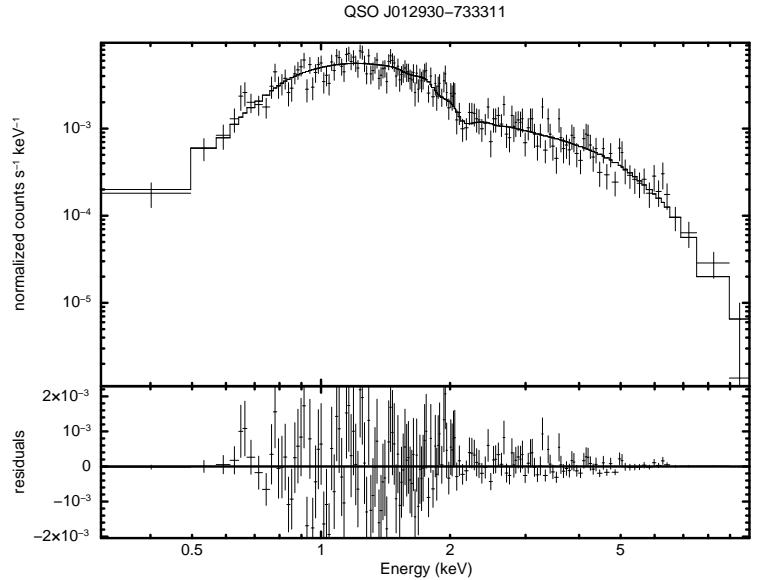}
\caption{{\em Upper panel}: the \cxo\ ACIS-I spectrum of QSO\,J012930-733311  
with the best-fitting absorbed power-law model (solid lines, see Section~\ref{sec:qso}); 
{\em Lower panel:} residuals between the model and the observed spectrum. }
\label{fig:qsosp} 
\end{figure}

Source\,3 is the most luminous X-ray object in NGC\,602a, and
corresponds to the candidate YSO from \citet[][their
source~52]{Goul2007} and \citet[][their Y283]{Carlson2011}.  FLAMES
spectroscopy revealed this source to be a quasar (QSO)
at a redsfhit of $z$\,$=$\,2.438\,$\pm$\,0.003 (labelled as `QSO' in
Fig.\,\ref{fig:hstcon}).  Following the IAU recommendations on
nomenclature we designate this as QSO\,J012930-733311
($\alpha$\,$=$\,01${^{\rm h}}$\,29${^{\rm m}}$\,30\fs17;
$\delta$\,$=$\,$-$73${^\circ}$\,33${'}$\,10\farcs77; J2000).  The
X-ray spectrum of the QSO can be well fitted by an absorbed power-law
model with $\Gamma$\,$\approx$\,1.8, and its observed flux is
1.1\,$\times$\,10$^{-13}$\,erg\,s$^{-1}$\,cm$^{-2}$. Its spectrum and
the best-fit model are shown in Fig.\,\ref{fig:qsosp}.

To model the absorbing column of gas towards QSO\,J012930-733311 we
used two components to account for Galactic and local SMC
absorption. The X-ray spectrum of QSO\,J012930-733311 does not show
apparent signs of the intrinsic absorption by the QSO, as well as any
cosmological absorption \citep[e.g.,][]{Behar2011}, so these components 
were ignored. Assuming solar abundances, the Galactic foreground absorption 
was fixed at a column density of 6\,$\times$\,10$^{20}$\,cm\,$^{-2}$
\citep{Wilms2000}.  A second component (with SMC abundances) was
fitted, yielding a value of (2.0\,$\pm$\,0.4)\,$\times$\,10$^{21}$\,cm$^{-2}$. 

This value compares well with the SMC neutral hydrogen density toward 
NGC\,602a of (2.0\,--\,2.5)\,$\times$\,10$^{21}$\,cm$^{-2}$ from the radio maps
from \citet{Stanimir1999}, which are sensitive to all angular
(spatial) scales between $98''$ (26\,pc) and $4\degr$ (4\,kpc). 

The reddening toward Sk\,183 was estimated by \citet{Evans2012} to 
be E$(B-V)$\,$=$\,0.09\,mag.  This corresponds to 
$\approx 4\times 10^{21}$\,cm$^{-2}$ \citep{Bouchet1985}.

We therefore have at least three independent estimates -- in good
agreement -- of the absorbing column towards \ngc, i.e., from fits to the
X-ray spectrum of QSO\,J012930-733311, from the H\,{\sc i} maps of the
SMC \citep{Stanimir1999}, and from optical photometry/spectroscopy of the
massive stars in the cluster.

\subsection{Source 4}
\label{sec:sgd12}
Source 4 is 1\farcs04 ($\sim$\,0.3\,pc) from star HTCP\,2
\citep{Hutchings1991}, which they classified as O9. HTCP\,2 was later
resolved into a close pair of stars by the {\em HST} imaging (see
Fig.\,\ref{fig:ob}): SGD\,5 and 12, with 15.371\,mag. and 15.816\,mag.
in the F555W filter respectively \citep{Schmalzl2008}. The FLAMES--Medusa 
fibers subtend 1\farcs2 on the sky, so our spectroscopy of HTCP\,2 will include
contributions from both SGD\,5 and 12.  Within the separate spectra we
saw no qualitative evidence for radial velocity shifts, and the
combined spectrum was classified as O9.5~V. The astrometric
uncertainty on the position of the X-ray source is 0\farcs5 (see
Table\,\ref{tab:ngc602src}), thus we can not securely identify SGD\,5
and/or 12 as counterparts.  Nevertheless, we note that source~4 has a
modest X-ray count rate and is detected in a soft band, which would be
expected from a normal OB-type star.

\subsection{Source 5}
\label{sec:sgd13}
\begin{figure}
\centering
\includegraphics[width=0.9\columnwidth]{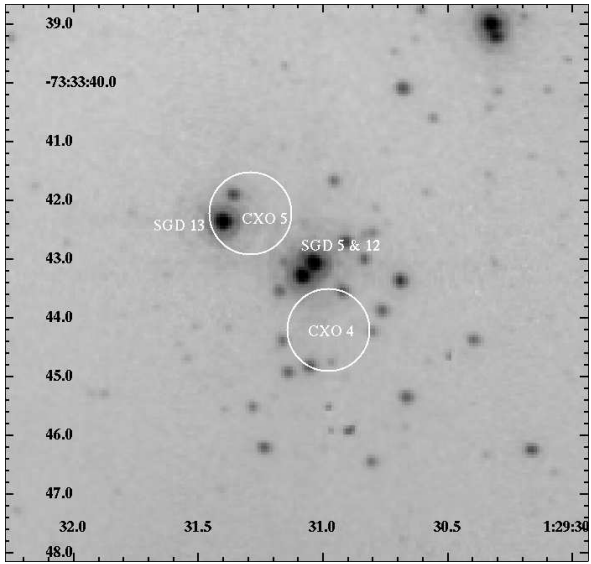}
\caption{{\em HST} F658N image of the central
  $\sim$\,10$''$\,$\times$\,10$''$ ($2.6\,{\rm pc}\times 2.6$\,pc) of \ngc.  Bright OB-type
  stars are labelled with identifications from
  \citet[][SGD]{Schmalzl2008}. Point sources detected in the \cxo\
  observations (see Table\,\ref{tab:ngc602src}) are shown by the white
  circles (with a radius of $0\farcs7$).}
\label{fig:ob} 
\end{figure}

There are two sources within the astrometric uncertainty of the X-ray
position for source~5 (see Fig.\,\ref{fig:ob}).  Both potential
counterparts have photometry from \citet{Schmalzl2008}: 
SGD\,13 has 15.663\,mag. in F555W filter and 15.871\,mag. in F814W filter; 
SGD\,119 has 19.201\,mag. in F555W filter and 19.228\,mag. in F814W filter.
 We have an optical spectrum of SGD\,13 from FLAMES, but
it was only observed with the Giraffe LR02 setting (covering
$\lambda\lambda$3960-4564\,\AA).  We provisionally classify the
spectrum as O9.5:~V, but note that we lack full coverage of the
optical region usually employed for classification of massive stars.
Although SGD\,13 and 119 are only separated by $\sim$0\farcs5, their
relative magnitudes mean that SGD\,13 should dominate the spectrum.

To investigate the nature of the fainter source, we take the {\em HST}
filters as approximately $V$- and $I$-bands, and adopt the intrinsic
colour for an O9.5 dwarf from \citet{Johnson1966} and extinction
relations from \citet{Howarth1983}. The approximate line-of-sight
reddening toward SGD\,13 from these assumptions is
E$(B-V)$\,$\approx$\,0.16\,mag.  Adopting the same reddening for
SGD\,119 to estimate its absolute magnitude and intrinsic colour
suggests it as a mid-late B-type dwarf. We note that if SGD\,119 were
a magnetic B-type star, it is unlikely that its X-ray luminosity would
be as large as that detected for source~5 \citep[cf.][]{Oskinova2011}.

Provided SGD\,119 is not a background source, SGD\,13 seems the most
likely counterpart, but given the limited wavelength coverage of the
FLAMES spectrum it is hard to speculate further on its nature at this
point.  If SGD\,13 were the genuine counterpart, its X-ray luminosity
($L_{\rm X}\,=\,3\pm 1\times 10^{32}$\,erg\,s$^{-1}$) would
correspond to log\,$L_{\rm X}$/$L_{\rm bol}$\,$\lsim$\,$-$6.3.
Following the same arguments as in Section~\ref{sec:mas}, a
colliding-wind system could account for the larger-than-expected X-ray
luminosity. Equally, the object could be analogous to the X-ray
variable system $\theta^2$\,Ori\,A, which has a quiescent X-ray
luminosity of $\sim$\,8\,$\times$\,10$^{31}$\,erg\,s$^{-1}$
(D.\,Huenemoerder, private communication), with its behaviour
suggested by \citet{Mitschang2011} as comparable to what would be
expected from a magnetically-confined wind model.

\subsection{Source 6} 

Source 6 is coincident with the candidate YSO `Y358' from
\citet{Carlson2011}. At the distance of \ngc, its X-ray luminosity
would be $L_{\rm
  X}$\,$\approx$\,6\,$\times$\,10$^{32}$\,erg\,s$^{-1}$, which is
rather high for a low-mass YSO. The optical magnitudes of Y358 are too
faint for it to be a young massive star in this region of relatively
low extinction \citep[22.695\,mag. in F555W filter and
21.970\,mag. in F814W filter][]{Schmalzl2008}. Given its relative isolation, we
suggest that source 6 is a background object rather than a genuine
YSO.
 
\subsection{Source 7}
Source 7 coincides with a spiral galaxy \citep[`G372'
from][]{Carlson2011}, which can be seen clearly in the {\em HST} image
(Fig.\,\ref{fig:hstcon}).

\subsection{Source 9}
Source 9 is $\approx 1\farcs6$ away from an optical source SGD~3593  
\citep[F555W = 25.892\,mag. and F814W = 24.809\,mag.][]{Schmalzl2008}.
Visual inspection of the {\em HST} ACS images shows that the optical 
object is fuzzy and strongly resembles a galaxy. We thus believe that
\cxo\ source 9 is a background galaxy.

\bibliographystyle{apj}
\bibliography{cluster}

\end{document}